\documentclass[
]{ceurart}

\usepackage{listings}
\usepackage{rotating}
\usepackage{graphicx}
\usepackage{subcaption}
\usepackage{booktabs}
\usepackage{graphicx}
\usepackage{pdflscape}   
\usepackage{adjustbox}   

\definecolor{interfcol}{RGB}{180,0,120}    
\definecolor{seccol}{RGB}{200,100,0}       
\definecolor{sneakcol}{RGB}{0,130,180}     
\definecolor{obstcol}{RGB}{0,140,80}       

\begin{document}

\copyrightyear{2022}
\copyrightclause{Copyright for this paper by its authors.
  Use permitted under Creative Commons License Attribution 4.0
  International (CC BY 4.0).}

\conference{ Bridge
 Over Troubled Water: Aligning Commercial Incentives With Ethical Design Practice To Combat Deceptive Patterns. Workshop at the 2026 CHI Conference on Human Factors in Computing Systems (CHI EA '26), April 13–17, 2026, Barcelona, Spain. }

\title{Deception by Design: A Temporal Dark Patterns Audit of McDonald's Self-Ordering Kiosk Flow}

\tnotemark[1]

 \author[1]{Aditya Kumar Purohit}[%
 orcid=0000-0002-9766-6575,
 email=aditya.purohit@cais-research.de
 ]

 \address[1]{Center for Advanced Internet Studies (CAIS), Konrad-Zuse-Strasse 2a, 44801 Bochum,Germany}

 \author[2]{Yuwei Liu}[%
 orcid=0009-0007-6035-2111,
 email=yuwei.liu@unine.ch,
 ]

 \address[2]{University of Neuchâtel, Av. du Premier-Mars 26, 2000 Neuchâtel, Switzerland}

 \author[2]{Manon Berney}[%
 orcid=0000-0002-7241-0127,
 email=manon.berney@unine.ch,
 ]

 \author[1]{Hendrik Heuer}[%
 orcid=0000-0003-1919-9016,
 email=hendrik.heuer@cais-research.de,
 ]

 \author[2]{Adrian Holzer}[%
 orcid=0000-0001-7946-1552,
 email=adrian.holzer@unine.ch,
 ]


\begin{abstract}
Self-ordering kiosks (SOKs) are widely deployed in fast food restaurants, transforming food ordering into digitally mediated, self-navigated interactions. While these systems enhance efficiency and average order value, they also create opportunities for manipulative interface design practices known as dark patterns. This paper presents a structured audit of the McDonald's self-ordering kiosk in Germany using the Temporal Analysis of Dark Patterns (TADP) framework. Through a scenario-based walkthrough simulating a time-pressured user, we reconstructed and analyzed 12 interface steps across intra-page, inter-page, and system levels. We identify recurring high-level strategies implemented through meso-level patterns such as adding steps, false hierarchy, bad defaults, hiding information, and pressured selling, and low-level patterns including visual prominence, confirmshaming, scarcity framing, feedforward ambiguity, emotional sensory manipulation, and partitioned pricing. Our findings demonstrate how these patterns accumulate across the interaction flow and may be amplified by the kiosk's linear task structure and physical context. These findings suggest that hybrid physical–digital consumer interfaces warrant closer scrutiny within emerging regulatory discussions on dark patterns.
\end{abstract}



\maketitle

\section{Introduction and Related Work}

Self-ordering kiosks (SOKs) have become a dominant ordering modality in quick-service restaurants (QSRs), with McDonald's operating more than 130,000 kiosks globally~\cite{rbr2024kiosks}. By enabling customers to browse menus, customize items, make payments, and complete transactions independently, SOKs transform food ordering into a digitally mediated, self-navigated experience \cite{lee2025design}.

Prior to widespread SOK deployment, nudges in offline food ordering environments were commonly used to promote healthier and more sustainable choices \cite{pandey2025nudging, langen2022nudges, lemken2023public}. Restaurants and cafeterias have employed strategies such as product placement, pricing structures, framing, and convenience-enhancing arrangements to increase the accessibility and attractiveness of selected options \cite{ensaff2021nudge, harbers2020effects, broers2017systematic, kraak2017novel}. While such nudges are often justified as public health interventions, the same choice architecture principles can equally serve to promote higher-margin or energy-dense products.

As SOKs have extended these choice architecture principles into digital interfaces, public debates have raised concerns about manipulative design practices that steer consumers toward economically or nutritionally suboptimal choices \cite{zardotab2024kiosks, mileyehh2021mcdonalds}. These practices are commonly described as \textit{dark patterns}, defined as interface design strategies that benefit service providers at the expense of user autonomy or informed decision-making \cite{mathur2021makes}. Industry reports suggest that SOKs increase average order values by 11--30\% at McDonald's \cite{mccauley2018measuring}, with executives attributing this to extended browsing time and enhanced digital prompts \cite{whitten2018mcdonalds}.

Although the concept of dark patterns was introduced by Brignull in 2010 \cite{brignull2018darkpatterns}, HCI research has primarily examined their manifestation in websites and mobile applications across domains such as e-commerce, social media, and gaming \cite{karagoel2021dark, gunawan2021comparative}. The research is scarce on investigating dark patterns in hybrid physical-digital environments such as QSR kiosks, a gap that carries real consequences given the financial, nutritional, and environmental implications of over-ordering in fast food contexts.

To address this, we conduct a structured audit of the McDonald's SOK system in Germany using the Temporal Analysis of Dark Patterns (TADP) framework proposed by~\citet{TADP_Method}. We analyze 12 interface steps across intra-page, inter-page, and system levels to examine how dark patterns accumulate temporally along a high-velocity interaction path.

Our findings reveal recurring high-level strategies of obstruction, social engineering, sneaking, and interface interference. These are implemented through meso-level patterns such as adding steps, creating barriers, false hierarchy, bad defaults, hiding information, pressured selling, and choice overload, and realized at the low level through visual prominence, confirmshaming, scarcity framing, feedforward ambiguity, emotional sensory manipulation, and partitioned pricing.

These findings carry implications beyond interface evaluation. As dark patterns increasingly become a focus of EU consumer protection and digital regulatory frameworks, our results suggest that hybrid physical–digital systems such as self-ordering kiosks warrant closer scrutiny. In particular, temporally layered manipulation across interaction flows may challenge regulatory approaches that focus primarily on isolated interface elements in online environments.

The remainder of this paper is structured as follows. Section 2 presents the temporal analysis, and Section 3 discusses the limitations and future research directions.

\section{Temporal Analysis: McDonald's in Germany}
This section presents the detailed application of the TADP framework to a McDonald’s SOK in Germany as an illustrative case. 
We chose Germany as the study location for two reasons: 1) geographic proximity for data collection, and 2) it is an EU Member State operating under a consumer protection and digital regulatory framework shaped by the Unfair Commercial Practices Directive~\cite{EC2021UCPD} and the Digital Services Act~\cite{dsa2022}. We describe the case selection, data collection, and analysis procedure to map temporally layered dark patterns across intra-, inter- and system levels.

\subsection{Data Collection and Reconstruction}
Data were collected in situ by co-authors through direct interaction with the McDonald's kiosk. Multiple walkthroughs of the ordering process were conducted to fully capture the interface configuration. 
The interaction flow from the start screen to payment confirmation was recorded. 
Screens were extracted and reconstructed in Omnigraffle, where we annotated interface elements, interaction costs (e.g., number of taps required for acceptance vs. refusal), navigation dependencies, and potential dark pattern instances. 
This reconstruction served as an audit trail enabling systematic temporal mapping.

\subsection{Scenario Operationalization: Simulating a Hurried User}

To operationalize temporality under realistic constraints, we simulated a time-pressured, minimally deliberative user persona representing a hurried customer interacting in a public, queue-based environment. Reduced deliberation was operationalized as selecting the first salient or default option at each decision point, approximating low-reflection decision-making behavior. Rather than mapping all possible navigation branches, this approach traces how dark patterns accumulate along a typical high-velocity interaction path.

\subsection{Analytical Procedure}

Following the TADP~\cite{TADP_Method}, we analyzed the kiosk interface across three levels of temporality, tagging dark patterns based on the ontology proposed by \citet{Gray_2023}. At the intra-page level, we examine dark patterns occurring within a single screen; at the inter-page level, we analyze how patterns unfold across sequential screens and on the system level we analyse patterns consumers are exposed to at service delivery level.

\begin{figure}[htbp]
    \centering
    \includegraphics[width=0.8\linewidth, page=1]{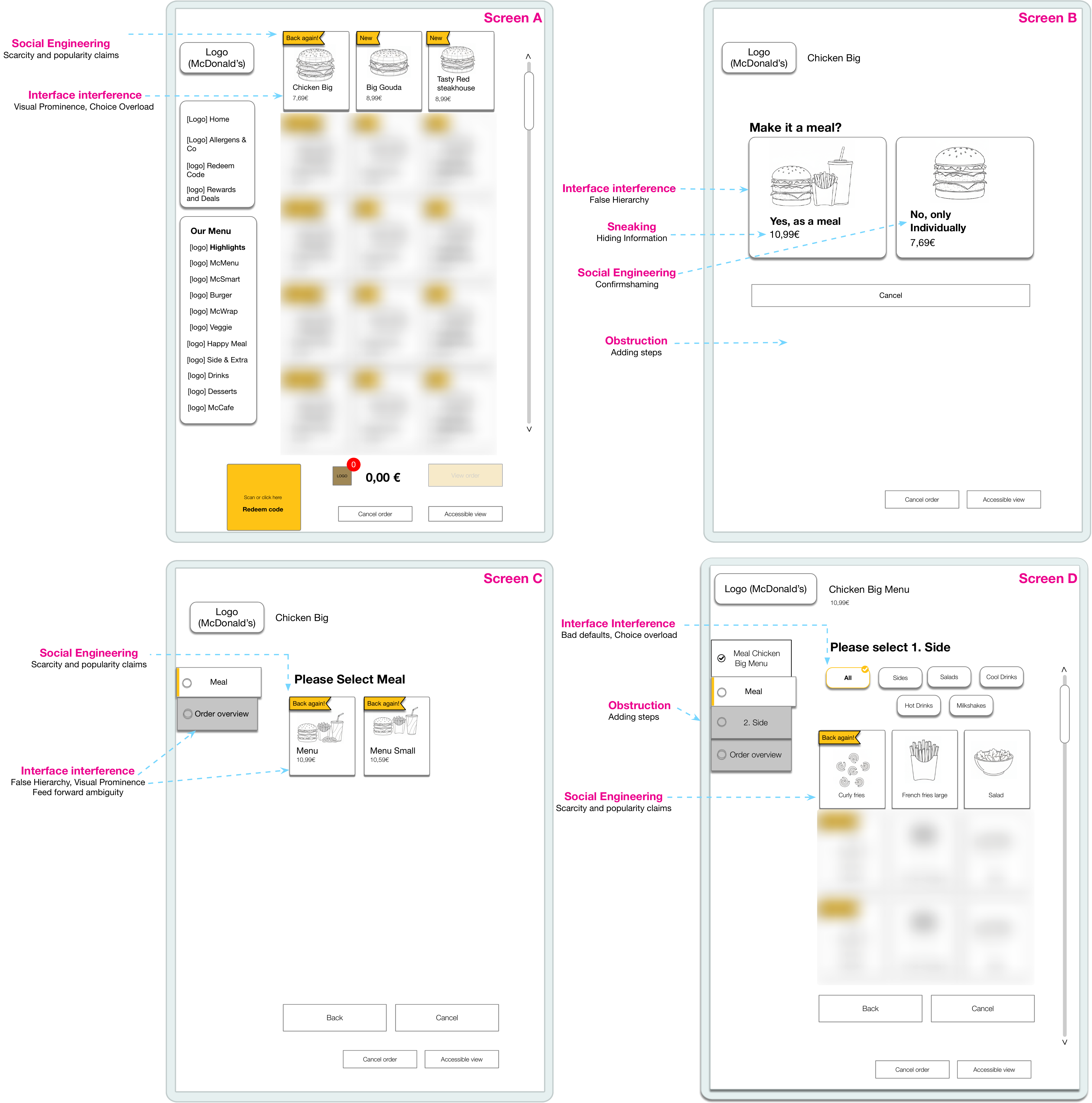}
    \caption{Intra-page analysis. Across screens A–D of the kiosk flow, annotations highlight social engineering, interface interference, sneaking, and obstruction within early product selection and meal upsell stages.}
    \label{fig:screen_atod}
\end{figure}

\textbf{\textit{Intra-page}}. As the hurried user interacts with the kiosk and clicks on ``highlights'' as the first salient option, they encounter screen A (see fig~\ref{fig:screen_atod}). This screen shows evidence of social engineering, a high-level dark pattern, implemented through the meso-level pattern of scarcity. Specifically, the design frames a returning item as a notable event, implying it was previously unavailable and may become unavailable again (scarcity), while also suggesting that high demand justified its return (social proof). A full-width banner in high-saturation yellow maximizes attentional capture, constituting a low-level pattern of visual prominence. A second meso-level pattern is also present on this screen. The highlights view presents 12+ items across four rows with no sub-categorization, forcing the user to evaluate options simultaneously. This cognitive overload is compounded by competing badges (New, Back Again!, Spicy, beef/veggie icons), which degrades autonomous decision-making and renders users more susceptible to visual hierarchy manipulations.

The Screen B on the top right in figure~\ref{fig:screen_atod} also employs several high-level patterns: interface interference, sneaking, social engineering, and obstruction. At the meso-level, manipulated choice architecture makes the meal option the first visually salient choice. At the low level, a confirmshaming pattern appears in the right-card label, where the word ``only'' frames the cheaper option as an inferior choice rather than a neutral economic preference. Finally, a classic drip pricing pattern is present where the bundled price is disclosed only at the upsell decision point, alongside the individual burger price. This mechanism leads the user to perceive the meal as a mere +3.30€ incremental cost once 7.69€ has been established as the baseline.

\begin{figure}[htbp]
    \centering
    \includegraphics[width=0.8\linewidth, page=1]{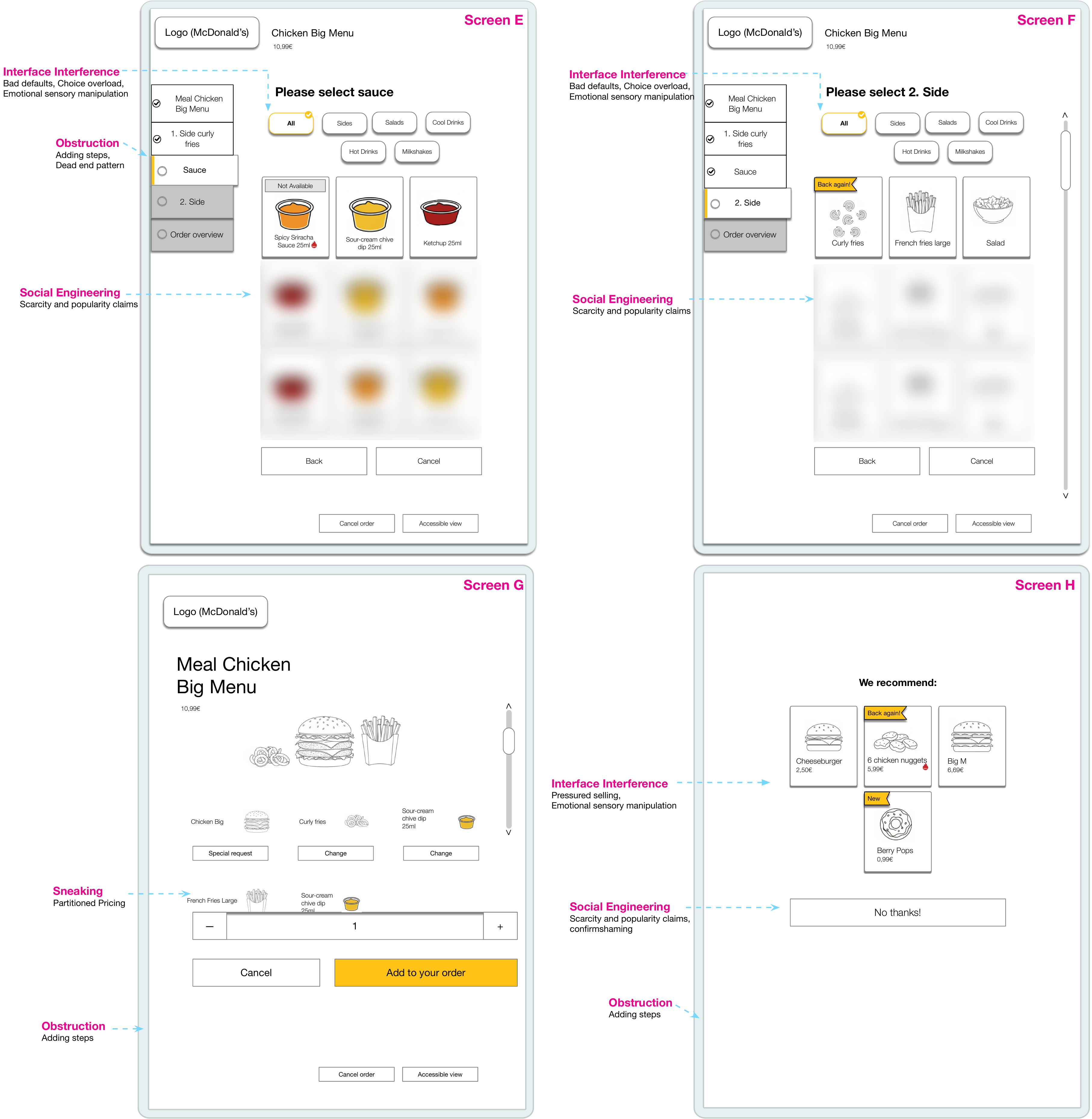}
    \caption{Continued intra-page layering of dark patterns across Screens E–H during meal configuration and cross-sell. Patterns include bad defaults, choice overload, partitioned pricing, pressured selling, and adding steps}
    \label{fig:screen_etoh}
\end{figure}

Similarly to screen B, the screen C in the lower left in figure~\ref{fig:screen_atod} again employs manipulated choice architecture, making the larger meal option the first visually salient choice. From this point onward, each screen is layered with dark patterns. Across Screens D, E, and F in fig~\ref{fig:screen_etoh} along with scarcity and bad defaults, a meso-level pattern of feedforward ambiguity appears in the left-side progress indicator, which fails to communicate how many steps remain before payment. These screens also exhibit the high-level pattern of obstruction in which the user cannot return to the home screen or skip directly to payment, forcing traversal through each upsell step. Screen G in fig~\ref{fig:screen_etoh}, the ``Add to Order'' screen, combines obstruction with a low-level pattern of partitioned pricing. Secondary side items (French Fries Large + Sour Cream-Chive Dip) appear as separate line items below the fold, requiring an active downward scroll to view the complete order. Costs and items are thus deliberately distributed across the visible and hidden areas of the screen.

\begin{figure}[htbp]
    \centering
    \includegraphics[width=0.8\linewidth, page=1]{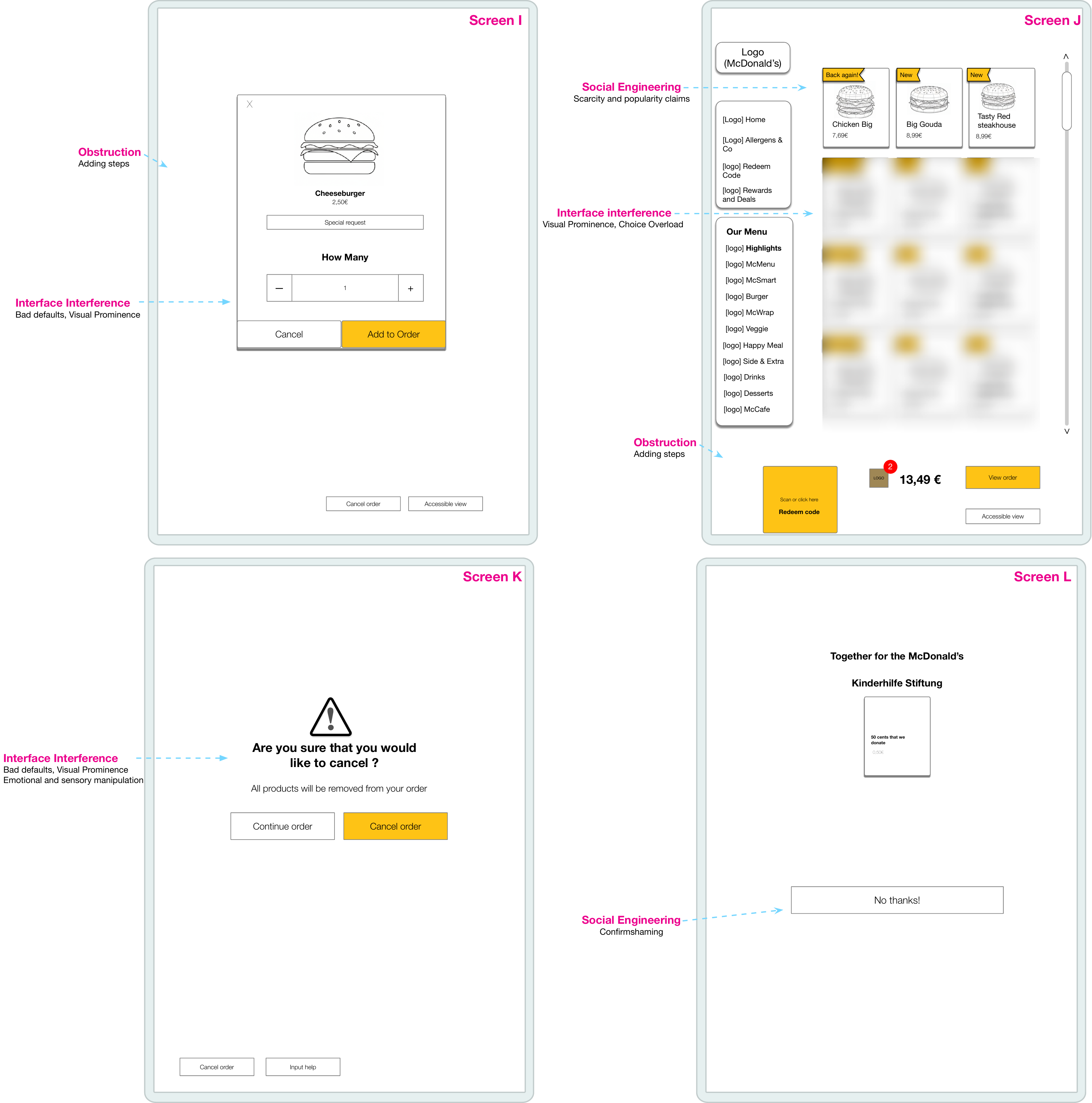}
    \caption{Late-stage dark patterns across Screens I–L, including upsell modals, cancellation prompts, and donation requests.
Obstruction, confirmshaming, visual prominence, and scarcity cues increase reversal costs.}
    \label{fig:screen_itol}
\end{figure}

Screen H employs a low-level pattern of pressured selling. A full-screen advertisement is injected between the completion of meal configuration and to view the order, presenting four additional items. This constitutes a second dedicated cross-sell screen (following the Cheeseburger modal in Screen I), and represents pressured selling through architectural commitment exploitation. The ``We recommend:'' label frames a profit-driven upsell as a personal suggestion from McDonald's, with no disclosed basis for why these specific items are recommended to this user. The dismiss button, reading ``No thanks!'', further reinforces this pattern. The exclamation mark and informal register frames the refusal as an emotionally loaded rejection rather than a neutral navigation choice.

\begin{sidewaysfigure}
  \centering
  \includegraphics[width=\textheight]{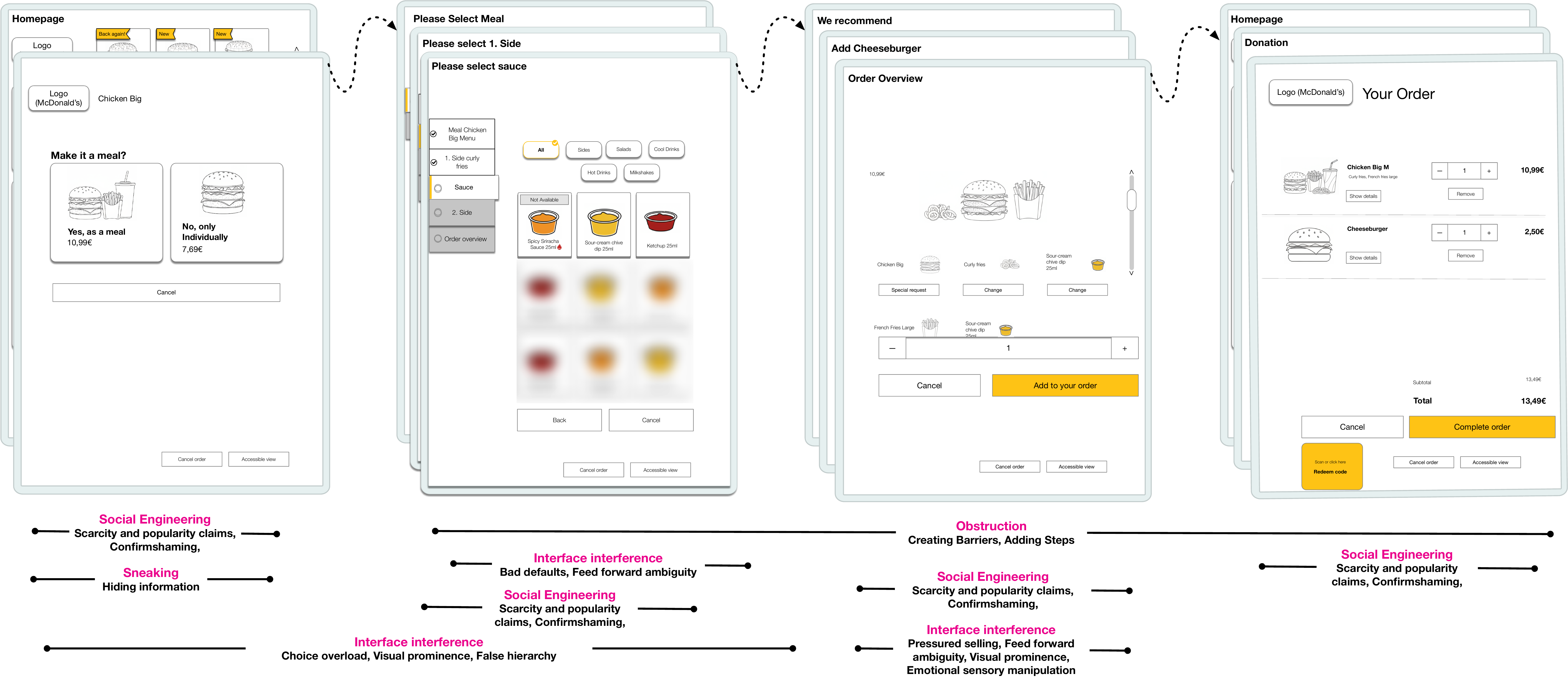}
  \caption{ Inter-page escalation of dark patterns across the 12-step kiosk flow. Recurring social engineering, interface interference, sneaking, and obstruction accumulate along the high-velocity interaction path.}
  \label{fig:inter-overview}
\end{sidewaysfigure}

\textbf{\textit{Inter-page}}. At the inter-page level, we examined how dark patterns unfold sequentially across Screens A–L (see fig~\ref{fig:screen_atod}, fig~\ref{fig:screen_etoh}, and fig~\ref{fig:screen_itol}) and whether they accumulate, repeat, or escalate over time (see figure~\ref{fig:inter-overview}). Rather than appearing as isolated interface elements, several patterns recur across consecutive steps, producing a layered escalation sequence along the high-velocity interaction path. As illustrated in figure~\ref{fig:inter-overview}, the 12-step kiosk flow reveals this escalation through the recurring presence of social engineering, interface interference, sneaking, and obstruction across consecutive stages.

Across the sequence, four high-level strategies recur: (1) Social Engineering, including scarcity cues, confirmshaming, and recommendation framing; (2) Interface Interference, including visual prominence, false hierarchy, choice overload, pressured selling, forward ambiguity, and bad defaults; (3) Obstruction, through added steps and prevention of direct progression to payment; and (4) Sneaking, via partitioned pricing and distributed cost visibility.

These strategies do not operate independently. They co-occur on multiple screens, producing layered manipulative effects. On Screen B (see figure~\ref{fig:screen_atod}), for instance, upsell screens combine false hierarchy with confirmshaming and obstruction through mandatory traversal. From a temporal perspective, the flow demonstrates progressive monetary escalation (Home $\rightarrow$ Burger $\rightarrow$ Meal $\rightarrow$ Side customization $\rightarrow$ Cross-sell $\rightarrow$ Additional modal $\rightarrow$ Donation $\rightarrow$ Home $\rightarrow$ Cart), with the interaction cost of refusal increasing at each step. Declining upsells requires repeated dismissals across multiple screens, and side selection traversal cannot be skipped once a meal is chosen (see Screen D in figure~\ref{fig:screen_atod}).

Even after reaching Screen G (see figure~\ref{fig:screen_etoh}), which provides an order overview, users cannot proceed directly to payment. Instead, they must navigate additional layers of interface interference, including pressured selling and social engineering (see Screen H in figure~\ref{fig:screen_etoh} and Screen I in figure~\ref{fig:screen_itol}), before being routed back to the homepage. There, meso-level strategies such as scarcity cues apply further upsell pressure, and payment is accessible only upon tapping the "view order" button on Screen J (see figure~\ref{fig:screen_itol}). Attempting cancellation triggers an additional confirmation dialog (see Screen K in figure~\ref{fig:screen_itol}), adding friction to reversal. This asymmetry between acceptance and decline paths aligns with TADP's emphasis on temporally layered manipulation, where cumulative exposure and repetition increase compliance likelihood under reduced deliberation.

\textbf{\textit{System level}}. At the system level, we analyze how the broader kiosk architecture and physical service environment can support or even amplify temporally layered dark patterns. While the primary focus is on interface screens, the self-ordering kiosk operates in a hybrid physical-digital environment that introduces additional contextual constraints. The kiosk is situated in a queue-based, public setting where users interact under time pressure, social visibility, and ambient noise. These conditions can reduce deliberation time and increase reliance on salient visual cues, potentially making interface interference strategies like visual prominence, false hierarchy, and bad defaults more effective than they would be in private digital contexts.

The kiosk employs a linear task architecture, requiring users to navigate through multiple configuration screens before reaching the payment section. This design, coupled with repeated upsell prompts, suggests that the obstruction is not confined to individual screens but is embedded within the overall flow structure. Hardware and interaction constraints further influence the user’s experience. Large touch targets, vertically stacked cards, and scroll-dependent visibility determine the order in which content appears. Additionally, partitioned pricing and below-the-fold item visibility hold significance when users need to inspect items while standing.

Finally, cancellation and reversal require additional confirmation steps, similar to the cancellation dialog (see screen k, fig~\ref{fig:screen_itol}), which increases friction for exiting the flow. Consequently, interaction costs are asymmetrically distributed at the architectural level rather than just isolated screens. Taken together, the system-level analysis suggests that dark patterns in the kiosk are not merely interface-level artifacts. They are supported by a broader service architecture that combines linear progression, mandatory traversal, environmental time pressure, and embodied interaction constraints. These contextual factors may amplify the cumulative effects identified in the inter-page analysis.

\section{Conclusion and Future Work}

This study presents a structured dark patterns audit of a McDonald's self-ordering kiosk (SOK) in Germany using the Temporal Analysis of Dark Patterns (TADP) framework. The analysis modeled a time-pressured user selecting the first salient option at each decision point, allowing us to examine how dark patterns accumulate along a high-velocity interaction path. While we do not claim generalizability across all kiosk configurations or user types, this work offers, to our knowledge, the first exploratory application of TADP to hybrid physical-digital consumer interfaces and illustrates how temporally layered manipulative mechanisms may operate in embodied ordering environments.

Future research should extend this analysis by examining additional user personas and alternative navigation paths. Comparative studies across countries and quick-service restaurant (QSR) chains, such as Burger King or KFC, could further investigate how regulatory environments, linguistic framing, and interface configurations shape the manifestation of dark patterns. Right-to-left interface orientations in Arabic-speaking regions, for instance, may alter visual hierarchy and default prominence in ways that current frameworks do not adequately capture. Such cross-cultural perspectives would contribute to ongoing debates on persuasive design scalability and region-specific regulatory frameworks.

To further assess the financial and nutritional implications of kiosk choice architecture, future work will map price and calorie distributions of promoted items across locations. This extension may support interdisciplinary discussions on consumer protection, public health, and ethical interface design in digital food environments.

\section*{Acknowledgment}
This paper is partially funded by SNSF project POWERPOSE, grant number 10004764.

\bibliography{0-sample-ceur}

@misc{rbr2024kiosks,
  author       = {{RBR Data Services}},
  title        = {Global Demand for Self-Ordering Kiosks Continues to Soar},
  year         = {2024},
  month        = jan,
  day          = {10},
  howpublished = {Datos Insights},
  url          = {https://datos-insights.com/press-release/global-demand-for-self-ordering-kiosks-continues-to-soar/},
  note         = {Accessed: 20 February 2026}
}

@article{lee2025design,
title = {Design of interactive systems: Information visualization methods of self-service technology in fast food restaurants},
journal = {Computers in Human Behavior Reports},
volume = {17},
pages = {100585},
year = {2025},
issn = {2451-9588},
publisher={Elsevier},
doi = {https://doi.org/10.1016/j.chbr.2024.100585},
author = {Yi-Shan Lee and Szu-Chi Chen and Yunqian Zhan and Meng-Cong Zheng}
}

@article{pandey2025nudging,
title = {Nudging strategies to promote plant-based and sustainable food consumption in canteens},
journal = {Appetite},
volume = {207},
pages = {107874},
year = {2025},
issn = {0195-6663},
doi = {https://doi.org/10.1016/j.appet.2025.107874},
publisher={Elsevier},
author = {Sujita Pandey and Annemarie Olsen and Marianne Thomsen}
}

@article{langen2022nudges,
title = {Nudges for more sustainable food choices in the out-of-home catering sector applied in real-world labs},
journal = {Resources, Conservation and Recycling},
volume = {180},
pages = {106167},
year = {2022},
issn = {0921-3449},
doi = {https://doi.org/10.1016/j.resconrec.2022.106167},
author = {Nina Langen and Pascal Ohlhausen and Fara Steinmeier and Silke Friedrich and Tobias Engelmann and Melanie Speck and Kerstin Damerau and Katrin Bienge and Holger Rohn and Petra Teitscheid},
publisher={Elsevier}
}

@article{lemken2023public,
	author = {Lemken, Dominic and Wahnschafft, Simone and Eggers, Carolin},
	doi = {10.1186/s12889-023-17127-z},
	id = {Lemken2023},
	isbn = {1471-2458},
	journal = {BMC Public Health},
	number = {1},
	pages = {2311},
	title = {Public acceptance of default nudges to promote healthy and sustainable food choices},
	volume = {23},
    publisher={Springer},
	year = {2023}}

@article{ensaff2021nudge,
  author    = {Ensaff, Hannah},
  title     = {A nudge in the right direction: the role of food choice architecture in changing populations' diets},
  journal   = {The Proceedings of the Nutrition Society},
  year      = {2021},
  volume    = {80},
  number    = {2},
  pages     = {195--206},
  month     = may,
  doi       = {10.1017/S0029665120007983},
  pmid      = {33446288},
  publisher={Cambridge University Press}
}

@article{harbers2020effects,
author = {Harbers, Marjolein C. and Beulens, Joline W. J. and Rutters, Femke and de Boer, Femke and Gillebaart, Marleen and Sluijs, Ivonne and van der Schouw, Yvonne T.},
	doi = {10.1186/s12937-020-00623-y},
	id = {Harbers2020},
	isbn = {1475-2891},
	journal = {Nutrition Journal},
	number = {1},
	pages = {103},
	title = {The effects of nudges on purchases, food choice, and energy intake or content of purchases in real-life food purchasing environments: a systematic review and evidence synthesis},
	volume = {19},
	year = {2020}}

@article{broers2017systematic,
    author = {Broers, Valérie J. V. and De Breucker, Céline and Van den Broucke, Stephan and Luminet, Olivier},
    title = {A systematic review and meta-analysis of the effectiveness of nudging to increase fruit and vegetable choice},
    journal = {European Journal of Public Health},
    volume = {27},
    number = {5},
    pages = {912-920},
    year = {2017},
    month = {06},
    issn = {1101-1262},
    doi = {10.1093/eurpub/ckx085}
}

@article{kraak2017novel,
author = {Kraak, V. I. and Englund, T. and Misyak, S. and Serrano, E. L.},
title = {A novel marketing mix and choice architecture framework to nudge restaurant customers toward healthy food environments to reduce obesity in the United States},
journal = {Obesity Reviews},
volume = {18},
number = {8},
pages = {852-868},
doi = {https://doi.org/10.1111/obr.12553},
year = {2017}
}

@misc{zardotab2024kiosks,
  author = {Zardotab},
  title = {Are those fast food kiosks lousy because they use "dark patterns" to up-sale?},
  howpublished = {Reddit},
  year = {2024},
  month = {March},
  url = {https://www.reddit.com/r/AskMarketing/comments/1b8kqgn/are_those_fast_food_kiosks_lousy_because_they_use/},
  note = {r/AskMarketing}
}

@misc{mileyehh2021mcdonalds,
  author = {Mileyehh},
  title = {The UX from the McDonald's self service kiosks is terrible},
  year = {2021},
  howpublished = {Reddit},
  note = {r/userexperience},
  url = {https://www.reddit.com/r/userexperience/comments/qjxico/the_ux_from_the_mcdonalds_self_service_kiosks_is/}
}

@inproceedings{mathur2021makes,
author = {Mathur, Arunesh and Kshirsagar, Mihir and Mayer, Jonathan},
title = {What Makes a Dark Pattern... Dark? Design Attributes, Normative Considerations, and Measurement Methods},
year = {2021},
isbn = {9781450380966},
publisher = {Association for Computing Machinery},
address = {New York, NY, USA},
doi = {10.1145/3411764.3445610},
booktitle = {Proceedings of the 2021 CHI Conference on Human Factors in Computing Systems},
articleno = {360},
numpages = {18},
location = {Yokohama, Japan},
series = {CHI '21}
}

@techreport{mccauley2018measuring,
  title        = {Measuring the {ROI} and Staying Power of a Self-Service Kiosk Restaurant Deployment},
  author       = {McCauley, Mary},
  institution  = {ZIVELO},
  year         = {2018},
  month        = may,
  type         = {White Paper},
  url          = {https://marketscale.com/wp-content/uploads/2018/10/ZIVELO_QSR_ROI_WhitePaper_FINAL-1-ilovepdf-compressed.pdf},
  address      = {Scottsdale, Arizona}
}

@misc{whitten2018mcdonalds,
  author = {Whitten, Sarah},
  title = {{McDonald's to add self-order kiosks to 1,000 stores each quarter}},
  journal = {CNBC},
  year = {2018},
  month = {June},
  day = {4},
  url = {https://www.cnbc.com/2018/06/04/mcdonalds-to-add-self-order-kiosks-to-1000-stores-each-quarter.html},
  urldate = {2026-02-17},
  note = {Interview with Steve Easterbrook on Squawk on the Street}
}

@misc{brignull2018darkpatterns,
  author       = {Brignull, Harry},
  title        = {Dark Patterns},
  year         = {2018},
  howpublished = {\url{https://darkpatterns.org/}},
  note         = {Accessed: 2026-02-16}
}

@article{karagoel2021dark,
author = {Ilayda Karagoel and Dan Nathan-Roberts},
title ={Dark Patterns: Social Media, Gaming, and E-Commerce},
journal = {Proceedings of the Human Factors and Ergonomics Society Annual Meeting},
volume = {65},
number = {1},
pages = {752-756},
year = {2021},
doi = {10.1177/1071181321651317}
}

@article{gunawan2021comparative,
author = {Gunawan, Johanna and Pradeep, Amogh and Choffnes, David and Hartzog, Woodrow and Wilson, Christo},
title = {A Comparative Study of Dark Patterns Across Web and Mobile Modalities},
year = {2021},
issue_date = {October 2021},
publisher = {Association for Computing Machinery},
address = {New York, NY, USA},
volume = {5},
number = {CSCW2},
doi = {10.1145/3479521},
journal = {Proc. ACM Hum.-Comput. Interact.},
month = oct,
articleno = {377},
numpages = {29}
}

@inproceedings{TADP_Method,
author = {Gray, Colin M. and Mildner, Thomas and Gairola, Ritika},
title = {Getting Trapped in Amazon's "Iliad Flow": A Foundation for the Temporal Analysis of Dark Patterns},
year = {2025},
isbn = {9798400713941},
publisher = {Association for Computing Machinery},
address = {New York, NY, USA},
doi = {10.1145/3706598.3713828},
booktitle = {Proceedings of the 2025 CHI Conference on Human Factors in Computing Systems},
articleno = {225},
numpages = {10},
location = {
},
series = {CHI '25}
}

@techreport{EC2021UCPD,
  author      = {{European Commission}},
  title       = {Guidance on the Interpretation and Application of Directive 2005/29/EC of the European Parliament and of the Council Concerning Unfair Business-to-Consumer Commercial Practices in the Internal Market},
  institution = {European Commission},
  year        = {2021},
  type        = {Commission Notice},
  number      = {2021/C 526/01},
  journal     = {Official Journal of the European Union},
  volume      = {C 526},
  pages       = {1--132},
  date        = {2021-12-29},
  note        = {Text with EEA relevance}
}

@legislation{dsa2022,
  title        = {Regulation ({EU}) 2022/2065 of the {European Parliament} 
                  and of the {Council} of 19 {October} 2022 on a Single Market 
                  For Digital Services and amending {Directive} 2000/31/{EC} 
                  ({Digital Services Act})},
  author       = {{European Parliament} and {Council of the European Union}},
  journal      = {Official Journal of the European Union},
  volume       = {L 277},
  pages        = {1--102},
  year         = {2022},
  month        = oct,
  day          = {27},
  number       = {2022/2065},
  url          = {https://eur-lex.europa.eu/legal-content/EN/TXT/?uri=CELEX%3A32022R2065},
  note         = {Text with EEA relevance}
}

@inproceedings{Gray_2023, series={CHI '23}, title={Towards a Preliminary Ontology of Dark Patterns Knowledge}, DOI={10.1145/3544549.3585676}, booktitle={Extended Abstracts of the 2023 CHI Conference on Human Factors in Computing Systems}, publisher={ACM}, author={Gray, Colin M. and Santos, Cristiana and Bielova, Nataliia}, year={2023}, month=apr, pages={1–9}, collection={CHI '23} }

\end{document}